\documentclass{article}
\usepackage[english]{babel}
\usepackage[utf8]{inputenc}
\usepackage[a4paper, total={6.5in, 9.5in}]{geometry}
\usepackage{graphicx}
\usepackage{booktabs,siunitx}
\usepackage{amssymb}
\usepackage{amsmath}
\usepackage{subcaption}
\usepackage{pgf,tikz}
\usepackage{listings}
\usepackage{hyperref}
\title{Solving Dynamic Discrete Choice Models: Integrated or Expected Value Function?}
\author{Patrick Kofod Mogensen}
\begin{document}
\maketitle
\section{Dynamic Discrete Choice Models}
Research in, and applications of, Dynamic Discrete Choice Models (DDCMs) has reached
some maturity. The history of the research program is well-established, and several
surveys exists, see for example \cite{rust1994structural} and \cite{aguirregabiria2010dynamic}.
In addition to surveys, the most relevant papers are \cite{rust1987optimal,john1988maximum,norets2010continuity}
for information on the \emph{expected value function}, and \cite{aguirregabiria2002swapping, aguirregabiria2010dynamic} for the
\emph{integrated value function} formulation. We briefly define relevant objects of interest,
as their definitions will be used below. The problem we're trying to solve is
\begin{align*}
    V(s) = \max_{a_0, \ldots a_T} E\left[\sum_{t=0}^T\left(\beta^{t}U(a_t,s_t))\right)|s_0\right]
\end{align*}
although we will consider the infinite horizon (stationary) dynamic programming formulation
\begin{equation}
    V(s) = \max_{a}\left(U(a,s)+\beta E_{s'}V(s')|s\right)
\end{equation}
for all purposes below. The choice set, $\mathcal{A}$, is discrete and
finite with $J$ elements. The first problem we face is, that the value function
depends on the shocks, and that these are typically continuous and unbounded -
and unobserved. Then, the usual theorems do not apply.

Below, we will use the the fact that the state $s$ consists of the \emph{observed}
states $x$ and the shocks $\epsilon$, such that $s=(x,\epsilon)$. We will also
assume Conditional Independence (CI), that is $(x, \epsilon)$ follows a (controlled)
Markov process according to the following transition density
\begin{align*}
    p(x',\epsilon'|x,\epsilon, i) = q(\epsilon'|x')p(x'|x,i).
\end{align*}
The shocks enter in a specific way
\begin{align*}
    U(s,a) = u(x, a)+\epsilon,
\end{align*}
that is they are additive.
\subsection{Expected Value Function (EV)}
The expected value formulation follows \cite{john1988maximum} and exploits the
results the author presented on the the continuity, differentiability,
and contraction properties of the expected value function in what \cite{aguirregabiria2010dynamic}
called Rust models. \cite{norets2010continuity} showed that the results extend beyond the restrictions
in the original formulation, but we will focus on \emph{Rust models} below. The
expected value function is \emph{defined} as
\begin{align*}
    EV_a(x) \equiv E_{s'}\left(V(s')|s,a\right) =  E_{s'}\left(V(x',\epsilon')|s\right)
\end{align*}
and leads to the following set of expected Bellman equations
\begin{align*}
    EV_a(x) &= E_{s'}\left(\max_{j}\left[U(j,x',\epsilon')+\beta EV_j(x')\right]|s,a\right),\quad\forall x\in \mathcal{X}, \forall a\in\mathcal{A}\\
    EV_a(x) &= \Gamma_{a}(EV,x).
\end{align*}
It solves two problems of handling $V(s)$ directly: the expected Bellman
operator is a contraction even if we have an unbounded (additive) component in
$U$, and we do not have to find the solution at each $\epsilon$, only each $x$.
If we assume extreme value type I shocks, we can further simplify the expression:
\begin{align*}
    \Gamma_{a}(EV,x) &= E_{x'}\left(\log\left[\sum_{j\in\mathcal{A}}\exp\left\{u(x, j)+\beta EV(x',j)\right\}\right]|x,a\right).
\end{align*}
We got rid of the numerical integration over shocks, and the maximization
step was replaced with the log-sum-exp calculation. The problem is now almost
a trivial exercise in root finding once we realize that Newton's method applied
to $\Gamma_{EV}(EV)-EV=0$ will give us convergence to some $\varepsilon$ tolerance
level in few iterations.
\subsubsection{Discrete $\mathcal{X}$}
If we assume that $\mathcal{X}$ is discrete and finite, we get
\begin{align*}
    \Gamma(EV, a) &= F(a)\left(\log\left[\sum_{j\in\mathcal{A}}\exp\left\{u(j)+\beta EV(j)\right\}\right]\right)
\end{align*}
where $F(a)$ is the conditional transition \emph{matrix} of dimension $|\mathcal{X}|\times|\mathcal{X}|$,
$\log$ and $\exp$ are applied element-wise, and $EV(j)$, $u(j)$ are (stacked over $x$) vectors of length $|\mathcal{X}|$.
\subsection{Integrated Value Function (W)}
We saw that \cite{john1988maximum} took the joint expectation over $s$ to get rid of
two problems: unboundedness of $U$, and $\epsilon$ as an explicit state variable.
Let us now consider an way of simplifying the problem. We consider the formulation
in \cite{aguirregabiria2002swapping}. They consider the \emph{integrated}
value function as the object of analysis, defined as
\begin{align*}
    W(x)\equiv E_{\epsilon'}\left[V(x,\epsilon')|\epsilon\right]
\end{align*}
leading to the integrated Bellman equations
\begin{align*}
    W(x) &= E_{\epsilon'}\left[\max_{j}\left\{U(j,x,\epsilon')+\beta E_{x'|x,j}W(x')\right\}|\epsilon\right],\quad\forall x\in \mathcal{X}\\
    W(x) &= \Lambda(W, x)
\end{align*}
which will solve the same problem as $\Gamma$. Again, we get the simplification
using extreme value type I shocks that
\begin{align*}
    \Lambda(W, x) = \log\left[\sum_{j\in\mathcal{A}}\exp\left\{u(x, j)+\beta E_{x'}(W(x'),|x,j)\right\}\right]
\end{align*}
Since we know that $\Gamma$ is
continuous and once differentiable, then so will $\Lambda$ be, as it is simply
a $\log$ to the sum of exponentials of the sum of a continuous function (utility) and
the EVs. They are also both contraction mappings with the same Lipschitz constant ($\beta$).
\subsubsection{Discrete $\mathcal{X}$}
Again, discrete and finite $\mathcal{X}$ simplifies matters, and we get
\begin{align*}
    \Lambda(W) = \log\left[\sum_{j\in\mathcal{A}}\exp\left\{u(j)+\beta F(a)W\right\}\right].
\end{align*}
Dimensions are as before, and all terms are stacked over $x$'s.
\section{Solution methods}
Since the comparison here is motivated by the solution part of the
nested fixed point algorithm in \cite{rust1987optimal}, we'll focus on two methods:
value function iterations and Newton steps for fixed point problems. We consider
discrete state spaces, although the lessons apply to the continuous state
case as well.

First consider the cost of \textbf{value function iterations} (VFIs). It is the same for $EV$ and
$V$. Consider again the expressions from above
\begin{align*}
    \Gamma(EV, a) &= F(a)\left(\log\left[\sum_{j\in\mathcal{A}}\exp\left\{u(j)+\beta EV(j)\right\}\right]\right)\quad\forall a\in\mathcal{A}\\
    \Lambda(W) &= \log\left[\sum_{j\in\mathcal{A}}\exp\left\{u(j)+\beta F(a)W\right\}\right].
\end{align*}
as long as we only calculate the log-sum-exp expression in $\Gamma$ once per major VFI
iteration, they both require $J$ matrix-vector multiplications, $J$ scalar-vector products,
$J$ vector-vector additions, $J$ $\exp$ calls over vectors and one $\log$ call over a vector.
This means that really is no reason to choose one over the other.

For \textbf{Newton's method} the computational costs differ. Define the conditional choice probabilities as
\begin{align*}
    P(a, x) &= Pr\left(a=\arg \max_{j\in\mathcal{A}} \{u(j, x)+\epsilon(j)+\beta E_{x'}(W(x')|x,j)\}\right)\\
    &= \frac{\exp\left(u(a,x)+\beta E_{x'}(W(x'|x,a)\right)}{\sum_{j\in\mathcal{A}}\exp\left(u(j,x)+\beta E_{x'}(W(x)|x,j)\right)}
\end{align*}
where the last line follows from the conditional independence and i.i.d. extreme value type I
assumptions from above. Then, stack the Bellman equations
\begin{align*}
    W &= \Gamma(W),\\
    EV &= \{\Gamma(EV, j)\}_{j\in\mathcal{A}}.
\end{align*}
Rearranging, we get
\begin{align*}
    W -\Gamma(W)&=(I -\Gamma)(W) =0,\\
    EV -\Lambda(EV &= (I-\Lambda)(EV) =0,
\end{align*}
where $I$'s denote the multiplicative identity, and $0$ is the zero element in
the Banach spaces where our value functions live. When
we say we are using Newton's method, it means applying Newton's method to these
equations by use of an inverse function theorem
\begin{align*}
W_{k+1} &\leftarrow W_{k} + (I-\Lambda')^{-1}(I-\Lambda)W_{k}\\
EV_{k+1} &\leftarrow EV_{k} + (I-\Gamma')^{-1}(I-\Gamma)EV_{k}
\end{align*}
Looking at the discrete versions of the formul\ae, and calculating the partial
derivative state-by-state, we see it is $\beta$ times the controlled transition.
That is, for $W$ we get
\begin{align*}
    \Lambda' &= \left(\sum_{i=1}^{J}\beta P(i)*F(i)\right)\\
\end{align*}
For the $EV$, the derivative has more elements, and can be calculated as
(defining $P$ as the stacked conditional choice probabilities $\{P(j)\}_{j\in\mathcal{A}}$:
\begin{align*}
    \Gamma' &= \beta P*^\top \begin{pmatrix}F(1) & F(1) &\cdots&F(1)\\
F(2) & F(2) &\cdots&F(2)\\
\vdots &\vdots &\ddots&\vdots\\
F(J) & F(J) &\cdots&F(J)\end{pmatrix}
\end{align*}
where $*^\top$ is the Hadamard product between the vector $P$ and the matrix, but applied
down through the rows instead of across the columns as is usually done, such that
\begin{align*}
p*^\top M = (p*M^\top)^\top
\end{align*}
for a vector $p$ and matrix $M$. Here we have the problem of using the EV formulation: if we want to use Newton's method
(and we really do), then this is now a $J|\mathcal{X}|\times J|\mathcal{X}|$ problem,
or $J^2$ as large as the integrated value function problem. Note, that in regenerative
models it is possible to use a trick in the style of the original \cite{rust1987optimal}.
However, it does require a little care when specifying the transitions, and indeed
there is a mistake in the original application, see the appendix for the details.
\subsection{Mathematical Programming with Equilibrium Constraints}
The solution of the types of model we've considered here, is typically used in
NFXP to estimate parameters in a given model based on some data we observe. This
is quite naturally formulated as a constrained maximum likelihood problem as mentioned
in \cite{RustManual}, but NFXP essentially substitutes the constraints into the
likelihood function since the model can be efficiently solved using the methods
presented here. The MPEC (Mathematical Programming with Equilibrium Constraints) method
in \cite{SuJudd} basically keeps the restrictions on the feasible $EV$'s (the ones
that solve $EV=\Gamma(EV)$) in the formulation as equality constraints. They
then solve the constrained maximum likelihood problem using an interior point
solver. Since each point in the state space has a constraint for each
choice, $EV_j(x_k) = \Gamma(EV, j, x_k)$, the considerations mentioned above for
Newton's method also applies here. A much more concise specification is available, by
simply considering constraints of the form $W(x_k) = \Lambda(W, x_k)$ instead.
\section{Numerical results}
We present some numerical evidence to support the claims above, and to
assess the practical importance of them. First, convergence is the same in the
two approaches. Consider \autoref{table:nk} for a sequence of the $\infty$-norm
of the differences between the current and previous iterates.
\begin{table}
    \centering
    \begin{tabular}{lll}
        \toprule
        k & W & EV\\
        \midrule
        1& $2.244 \times 10^1$ & $2.244 \times 10^1$\\
        2& $1.666 \times 10^1$ & $1.666 \times 10^1$\\
        3& $1.292 \times 10^{1}$ & $1.292 \times 10^{1}$\\
        4& $1.091 \times 10^{1}$ & $1.091 \times 10^{1}$\\
        5& $5.535 \times 10^{0}$& $5.531 \times 10^{0}$\\
        6& $2.883 \times 10^{0}$ & $2.879 \times 10^{0}$\\
        7& $3.635 \times 10^{-1}$& $0.363 \times 10^{-1}$\\
        8& $6.553 \times 10^{-3}$ & $0.007 \times 10^{-3}$\\
        9& $1.378 \times 10^{-6}$ & $1.362 \times 10^{-6}$\\
        10& $7.105 \times 10^{-14}$ & $6.928 \times 10^{-14}$\\
        \bottomrule
    \end{tabular}
    \caption{Convergence of Newton steps in Rust's model ($|\mathcal{X}|=1000$) measured in change in $EV$ for both cases.}
    \label{table:nk}
\end{table}
For a given $k$, we see the similar improvements.
What about differences in run-times?  In \autoref{table:times},
we see the time to perform a Newton step for the $EV$ formulation relative to
the $W$ formulation. We solve Rust's model, and what we call Hitsch's model.
The latter is a storable goods demand model. For a similar peer reviewed model
see \cite{hendel2006measuring}. This simpler version is due
to a presentation by G\"{u}nter J. Hitsch. \footnote{See the presentation titled
"Single Agent Dynamics:  Dynamic Discrete Choice" by G\"{u}nter J. Hitsch,
The University of Chicago Booth School of Business, 2013.} The $W$ formulation
is faster as expected. We also report the time to calculate the Newton step
and add it to the existing iterate. The "total" time also includes calculating
choice probabilities, the derivative of the Bellman operators, and so on.
\begin{table}
    \centering
\begin{tabular}{lllllll}
    Rust's model ($J=2$)      & &&  & \\
    \toprule
    $|\mathcal{X}|$& 10 & 100 & 200 & 300 & 500 & 800\\
\midrule
Total relative time (EV/W)     & 3.94 & 2.80 & 3.64 & 3.93 & 4.26 & 7.16\\
Only $x_{k-1}+(I-M)\backslash(x_k-x_{k-1})$    & 3.90 & 3.56 & 4.78 & 4.94 & 5.47 & 7.24 \\
\bottomrule\\
Hitsch's model ($J=3$)&&&\\
\toprule
$|\mathcal{X}|$     & 12  & 102 & 202  & 302 & 502 & 802 \\
\midrule
Total relative time (EV/W)   & 4.45 & 5.07 & 7.26 & 10.90 & 9.60 & 13.49 \\
Only $x_{k-1}+(I-M)\backslash(x_k-x_{k-1})$    & 3.93 & 8.40 & 13.10 & 16.70 & 15.84 & 16.56 \\
\bottomrule
\end{tabular}
\caption{Relative run-times of Newton-steps. Times are given as EV formulation relative to W formulation, both total time and the time it takes to calculate the step and take it.}
\label{table:times}
\end{table}
\section{Conclusion}
\cite{rust1987optimal,john1988maximum} introduced simplifying assumptions that facilitated
solution and estimation of empirically models of dynamic, discrete decision making.
The theory and analysis was based on the expected value function. As shown,
this is \emph{not} efficient if Newton's method is applied to solve the system.
The same information contained in the $EV(a)$'s can be summarized
by the lower dimensional $W$. If successive approximations are used, there is no
advantage, but if higher order methods are applied, it is inefficient to use $EV$
over $W$. This also applies to the MPEC approach in \cite{SuJudd}, which will have
more constraints than necessary. The performance gap increase as the number of
choices increases, so especially multinomial problem is many discrete choices
should be solved with $W$ instead of $EV$.
\appendix
\numberwithin{equation}{section}
\section{Appendix: Rust's Fr\'{e}chet derivative}
In \cite{rust1987optimal} we have a binary model, so the expected value function can be separated into
$EV_1(x)$ and $EV_2(x)$ in the notation of this note. In the original paper,
a convenient observation is made:
\begin{align*}
EV_2(x) = EV_2(0) = EV_1(0)
\end{align*}
that is, once Harold Zurcher replaces the engine it is as if the bus is brand new
and replacement was not chosen. This basically means that we can fully represent
the solution of the model using $EV_1(x)$, and then the Newton steps simplify to
(writing $EV \equiv EV_1$)
\begin{align*}
EV_{k+1} &\leftarrow EV_{k} + (I-\beta (P(1)*^\top F(1) +P(2)*^\top F(2) )^{-1}(I-\Gamma)EV_{k}
\end{align*}
This is a substantial dimensionality reduction, and eases the computational effort
as a result. If there are $|\mathcal{X}|$ discrete states, then $\Gamma'$ is only of size
$|\mathcal{X}|\times|\mathcal{X}|$ instead of $2|\mathcal{X}|\times2|\mathcal{X}|$.
This is computationally very convenient, and it contributes to the fast run times
in \cite{iskhakov2016comment}.

However, there is a slight logical disconnect between the way $EV$ is reduced to
effectively be of size $|\mathcal{X}|$ rather than $2|\mathcal{X}|$, and the
way the Fr\'{e}chet derivative of the Bellman operator is derived to be Rust (2000, p. 25):
\begin{align*}
    P(1)*^\top F(1)+\left[P(2)*^\top\begin{pmatrix}1 & 0 & \ldots & 0\\1 & 0 & \ldots & 0\\\vdots & \vdots& \ddots&\vdots\\1 & 0 & \ldots & 0\end{pmatrix}\right]
\end{align*}
where (choosing a small $p$ vector for demonstration purposes)
\begin{align*}
    F(1)=
    \begin{pmatrix}
     p_1  & p_2 & p_3 & 0  & 0\\
     0  & p_1 & p_2 & p_3  & 0\\
     0  & 0 & p_1 & p_2 & p_3\\
     0 & 0 & 0 & p_1  & p_2+p_3\\
     0 & 0 & 0 & 0  & 1\\
    \end{pmatrix}.
\end{align*}
The interpretation here is, that if you replace, you will reset your odometer,
but you will have to keep the bus in the garage for a month. The problem is, that
if that is the modeling choice, and the controlled process is as above, then
generally $EV_1(x)$ is \emph{not} $EV_2(0)$, as is clearly seen:
\begin{align*}
    EV_1(x) &= p_1W(x)+p_2W(x+\Delta x)+p_3W(x+2\Delta x)\\
    EV_2(x) &= 1*W(0)\\
            &= EV_2(0)
\end{align*}
where $\Delta x$ is the change in $x$ from one bin to the following. Now, it \emph{is}
true that $EV_2$ is constant, but as $W(x)$ is clearly going to be decreasing
due to rising maintenance costs, we have that $EV_1(x)<EV_2(x)$ as long as $p_1<1$.
This is unfortunate, as this increases the cost of the Newton steps substantially,
 if we want to keep the interpretation of the transitions.
The fix is easy, just specify $F(2)$ as
\begin{align*}
    \begin{pmatrix}p_1 & p_2 & p_3& 0 &\ldots & 0\\p_1 & p_2 & p_3& 0 &\ldots & 0\\p_1 & p_2 & p_3& 0 & \ldots & 0\\\vdots&\vdots&\vdots&\vdots&\ddots&\vdots \\p_1 & p_2 & p_3& 0& \ldots & 0\end{pmatrix}
\end{align*}
Then it is actually true, that, if we enter a period where the engine was replaced
last period, it is the same situation as entering a period with a bus that did
not have an engine replacement last period, but didn't transition from the first
to the second state either.

\bibliographystyle{plain}
\bibliography{ref}

\end{document}